\documentclass[journal]{IEEEtran}
\IEEEoverridecommandlockouts
\usepackage{cite}
\usepackage{graphicx,  algorithm,amsmath,amssymb,amsfonts,textcomp,xcolor,caption,subcaption}
\usepackage{algpseudocode}
\usepackage{hyperref}
\graphicspath{ {Figures/} }
\begin{document}

\title{\huge iRDRC: An Intelligent Real-time Dual-functional Radar-Communication System for Automotive Vehicles\\
}

\author{Nguyen Quang Hieu, Dinh Thai Hoang, \IEEEmembership{Member, IEEE}, Nguyen Cong Luong, and Dusit Niyato, \IEEEmembership{Fellow, IEEE} \\

\thanks{N. Q. Hieu and D. Niyato are with the School of Computer Science and Engineering, Nanyang Technological University, Sinapore  (e-mail: \{quanghieu.nguyen, dniyato\}@ntu.edu.sg.}
\thanks{D. T. Hoang is with the School of Electical and Data Engineerng, University of Technology Sydney, Sydney, NSW 2007, Australia (e-mail: hoang.dinh@uts.edu.au).}
\thanks{N. C. Luong is with the Faculty of Computer Science, PHENIKAA University, Hanoi 12116, Vietnam (e-mail:luong.nguyencong@phenikaa-uni.edu.vn).}
}
\maketitle

\begin{abstract}
This letter introduces an intelligent Real-time  Dual-functional Radar-Communication (iRDRC) system  for autonomous vehicles (AVs). This system enables an AV to perform both radar and data communications functions to maximize bandwidth utilization as well as significantly enhance safety. In particular, the data communications function allows the AV to transmit data, e.g., of current traffic, to edge computing systems and the radar function is used to enhance the reliability and reduce the collision risks of the AV, e.g., under bad weather  conditions. The problem of the iRDRC is to decide when to use the communication mode or the radar mode to maximize the data throughput while minimizing the miss detection probability of unexpected events given the uncertainty of surrounding environment. To solve the problem, we develop a deep reinforcement learning algorithm that allows the AV to quickly obtain the optimal policy without requiring any prior information about the environment. Simulation results show that the proposed scheme outperforms  baseline schemes in terms of data throughput, miss detection probability, and convergence rate. 
\end{abstract}


\section{Introduction}
\IEEEPARstart{A}UTONOMOUS vehicles (AVs) are required to navigate efficiently and safely in complex and uncontrolled environments~\cite{ma2019}. To meet these requirements, Dual-Functional Radar-Communication (DFRC) system design has been recently proposed as a promising technology for AVs. The DFRC allows an AV to jointly implement radar and communication functions. In particular, with the radar function, the AV is able to accurately detect the presence of distant objects or unexpected events even under the bad weather conditions and poor visibility. With the communication function, the AV can use communication channels to communicate with road-side units, base stations, and edge computing systems, e.g., by using vehicle-to-infrastructure (V2I) and vehicle-to-network (V2N), to facilitate intelligent road management, route selection, and data analysis~\cite{kumari2015, choi2016}. 

Since the DFRC system implements both radar and communications using a single hardware device, these functionalities share some  system resources such as antennas, spectrum, and power. As a result, one major problem of the AV is how to optimize the resource sharing between the radar function and  communication function. In particular, the problem of the AV is how to optimize the selection between the radar mode and communication mode. 

Recently, some resource sharing approaches have been proposed to solve the problem. In particular, the authors in~\cite{kumari2015} proposed to adopt the IEEE 802.11ad standard for the joint radar-communication in an AV system. Accordingly, the AV reserves  preamble blocks in the IEEE 802.11ad frame for the radar mode, i.e., to estimate their ranges and velocities, and uses data blocks for the data transmission. Different from~\cite{kumari2015}, the time sharing approach proposed in~\cite{chiriyath2017} uses time cycles instead of the standard frames. Then, time portions in the time cycle are allocated to the radar mode and communication mode to maximize the radar estimate rate and communication rate of the radar-communication system.
Consider the communication system for the AVs in the V2I scenario, the authors in~\cite{choi2016} proposed a method to reduce the beam alignment overhead between the AVs and infrastructures. However, the radar's performance on object detection is not considered.

In general, the approaches in~\cite{kumari2015, chiriyath2017, choi2016} are fixed schedule schemes that are not appropriate to implement in practice because the surrounding environment of the AV is uncertain and dynamic.
To maximize the resource efficiency under uncertain environment, adaptive algorithms for the radar and communication mode selection are required. For example, when the weather is in a bad condition, e.g., heavy rain, the AV can select the radar mode more frequently to improve the radar performance to detect unexpected events on the road. 
In contrast, when the weather and the communication channel are in good conditions, the AV can select the communication mode more frequently to transmit its data. 
However, it is challenging for the AV to determine optimal decisions because the environment states, e.g., weather and road states as well as the communication channel state are dynamic and uncertain. In this letter, we thus develop a deep reinforcement learning (DRL) technique that enables the AV to find the optimal selection of the radar mode and communication mode without prior knowledge of the environment. To the best of our knowledge, this is the first approach using DRL to solve the mode selection problem of the DFRC in AV.
For this, we first formulate the AV's problem as a Markov decision process (MDP). Then, we develop the DRL with Deep Q-Network (DQN)~\cite{mnih2015} algorithm to achieve the optimal policy for the AV. Simulation results show that the proposed DRL outperforms  baseline schemes in terms of higher data throughput, miss detection probability, and shorter convergence time. 
\section{System Model}
\begin{figure}[t]
\centering
\includegraphics[width=0.8\linewidth]{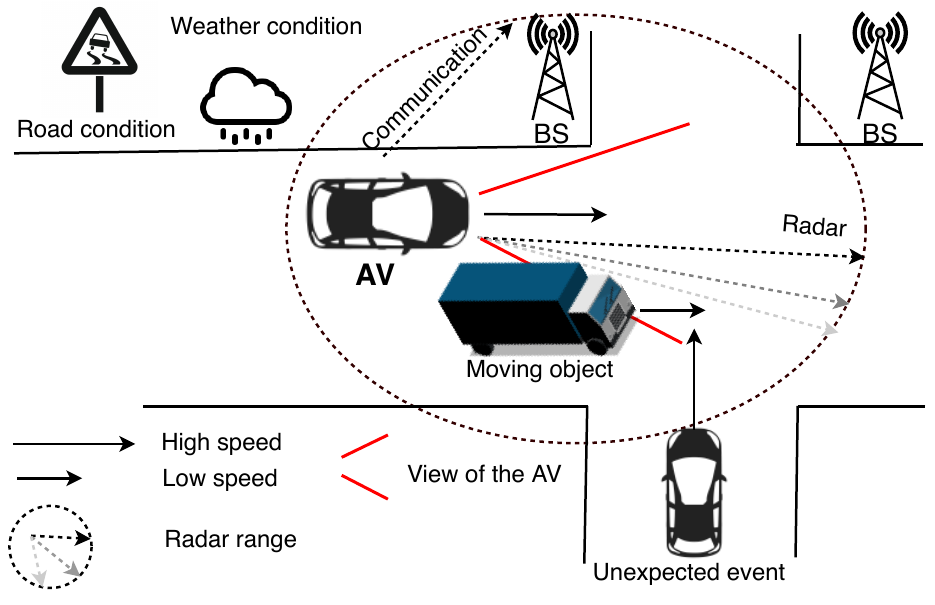}
 \caption{Autonomous vehicle with a DFRC system.}
  \label{AV-system-model}
  \vspace{-0.5cm}
\end{figure}

The system model with an AV as shown in Fig.~\ref{AV-system-model}. The AV is equipped with a DFRC equipment that enables the AV to work in two modes, i.e., the radar mode and the communication mode. Typically, the radar and communication modes can be allocated in time cycles, in which each time cycle is separated to radar mode and communication mode~\cite{chiriyath2017}. Unlike~\cite{chiriyath2017}, we consider that each time cycle/step is allocated to either radar mode or communication mode. This enables the AV to effectively change the mode based on the current observation of environment, rather than based on the previous time cycle as in~\cite{chiriyath2017}. 
\vspace{-0.3cm}
\subsection{Dual-functional Radar-Communication Model}
\label{subsec:dfrc-model}

In the communication mode, the AV uses the V2I capability to transmit the data, e.g., of current road traffic or live on-board video streaming, to the Base Stations (BSs) distributed along the road. Assume that the AV uses a single channel for the data transmission and has a data queue for storing incoming data packets, e.g., from its sensor devices. Let $D$ be the capacity of the data queue. In the radar mode, the AV performs an automotive millimeter-wave radar to detect unexpected events. As shown in Fig.~\ref{AV-system-model},  the radar mode can be used to detect unexpected events, e.g., a car coming from another road obscured by a truck.
In particular, we define an unexpected event as an event that can possibility cause collisions with the AV.
We consider that the occurrence of an unexpected event is influenced by four main factors: the road condition, weather condition, speed of the AV, and  nearby moving object~\cite{kloeden2001, ustransport}. Note that the values of these factors can be obtained by the AV's sensing system, e.g., road friction sensor, weather station instrument, speedometer, and cameras.

Let $r \in \{0,1\}$, $w\in \{0,1\}$, $v\in \{0,1\}$, and $m\in \{0,1\}$ be the road state, weather state, speed state, and moving object state, respectively. In particular,  $r=1$, $w=1$, $v=1$, and $m=1$ represent unfavorable conditions, e.g., slippery road, rainy weather, high speed of the AV, and a moving object nearby, respectively. In contrast, $r=0, w=0, v=0$ and $m=0$ express favorable conditions, e.g., straight road, good weather, low speed and without a moving object nearby, respectively. Let $p_j^i$ denote the probability to occur an unexpected event at the current condition $j$ (where $j \in \{0,1\}$ corresponds to favorable or unfavorable conditions, respectively) of factor $i$, $i \in \{r, w, v, m\}$.  For example, $p_1^r$ expresses the probability of an unexpected event to occur given the slippery road condition, i.e., $r=1$. 
Note that the generalization of the states beyond 0 and 1 is straightforward. For example, the speed of the AV can be divided into multiple levels, e.g., low, medium, and high.
\vspace{-0.3cm}
\subsection{Environment Model}

To model the dynamic of environment, the probabilities $p_j^v$, $p_j^w$ are taken from the real-world data in~\cite{kloeden2001, ustransport}, and other probabilities are assumed to be pre-defined. Then, we can determine the probability of an unexpected event to occur given factor states $(r,w,v,m)$ using the Bayes' theorem. For this, let $\oplus$ denote the occurrence of an unexpected event, and $\ominus$ denote that no unexpected event occurs.
Let $\tau_i$ be the probability that factor $i$ is at state $0$, where $i \in \{ r, w, v, m\}$. Thus, the probability that factor $i$ at state $1$ is $1 - \tau_i$. By using the Bayes' theorem, the probability of an unexpected event to occur given factor states $(r, w, v, m)$ is determined by:
\begin{equation}
P(\oplus) =  \sum_{i \in \{r, w, v, m\}} \left( \tau_i p_0^i + (1 - \tau_i) p_{1}^i \right).
\label{eq:unexpected-ev-prob}
\end{equation}

In general, when the probability of an unexpected event to occur, $P(\oplus)$,  is high, the environment is more dynamic and uncertain. 
We introduce a metric, i.e., the miss detection probability, to evaluate the performance of the proposed system. The miss detection probability is defined by the ratio of the number of unexpected events that the AV cannot detect to the total number of unexpected events on the road. A high miss detection probability results in a high risk of accident for the AV. We also introduce the second metric to evaluate the performance of the proposed system that is the data throughput. The data throughput is defined as the average number of packets per time unit that is successfully transmitted from the AV to the BSs. Note that, we assume that the accuracy of the autonomous radar system is perfect, i.e., there is no
miss detection or false alarm, when the AV uses the radar mode. However, the system model can be straightforwardly extended by considering the miss detection and false alarm caused by sensing accuracy of the radar. In this case, the proposed DRL scheme still can work well as it can learn these parameters through real-time interactions with the environment.

Intuitively, to minimize the miss detection probability, the AV can use the radar mode more frequently to detect unexpected events, but this reduces the data throughput. Conversely, to increase the throughput, the AV can use the communication mode more frequently, but this may increase the miss detection probability. Consider this tradeoff with the uncertainty of environment, the AV's decision making problem can be modeled as an MDP. We then develop a DRL algorithm to quickly obtain the optimal policy for the AV without requiring completed information about  environment. 
The details about the DRL scheme that enables the AV to quickly find the optimal policy will be discussed in Section~\ref{subsec:results}.

\section{Problem Formulation}
\label{sec:problem_formulation}
To formulate the problem by using the MDP, we define a tuple of $<\mathcal{S}, \mathcal{A}, \mathcal{R}, \mathcal{P}>$, where $\mathcal{S}$, $\mathcal{A}$, $\mathcal{R}$, and $\mathcal{P}$ are the state space, action space, reward function, and state transition probability of the AV, respectively. Note that the transition probability $\mathcal{P}$ is unknown to the AV in advance.
\subsection{Action Space and State Space}
At each time step, the AV decides to use either the communication mode or the radar mode. Let $\mathcal{A}$ denote the action space of the AV, $\mathcal{A} = \big\{a; a\in \left \{ 0,1 \right \} \big\}$, where $a=0$ means that the AV chooses the communication mode, and $a=1$ means that the AV chooses the radar mode.  
The state of the AV is the combination of (i) the state of the data queue, (ii) the state of the channel that the AV uses for its data communication, (iii) the state of the road, (iv) the weather state, (v) the speed state of the AV, and (vi) the nearby moving object state. Thus, the state space of the AV can be defined as 
\begin{eqnarray}
	{\mathcal{S}}	&=\Big\{ (d, c, r, w, v, m) ; d \in \{0,1,\ldots,D\}, c \in \{0,1\},  \nonumber \\ 
	& r \in \{0,1\}, w \in \{0,1\}, v \in \{0,1\}, m \in \{0,1\} \Big\},
\end{eqnarray}
where $d$ represents the state of the data queue, i.e., the number of packets in the data queue, $c$ refers to the state of the communication channel that the AV uses to transmit data to the BSs. $c=0$ if the channel is good, i.e., low interference, and $c=1$ if the channel is bad, i.e., high interference. $r, w, v$, and $m$ are defined in Section~\ref{subsec:dfrc-model}. The state of the system at time step $t$ is defined as $s_t = (d, c, r, w, v, m) \in \mathcal{S}$.
\vspace{-0.3cm}
\subsection{Reward Function}
At each time step $t$, the AV chooses an action $a_t \in \mathcal{A}$ at state $s_t \in \mathcal{S}$ and receives an immediate reward $r_t$. The reward is designed to encourage the AV to increase the data throughput and at the same time decrease its miss detection probability. For this, we define the reward function as follows.

When the AV selects the communication mode and if the channel state is good, the AV successfully transmits $\nu_1$ packets and receives a reward $r_1$. Otherwise, when the AV selects the communication mode and if the channel is bad, the AV successfully transmits $\nu_2$ packets and receives a reward $r_2$. Moreover, when the AV selects the communication mode and an unexpected event occurs, the AV receives a penalty of $-r_3$. When the AV selects the radar mode and if the AV does not detect any unexpected event, the AV receives no reward. Otherwise, when the AV selects the radar mode and if the AV detects an unexpected event, the AV receives a reward that is proportional to the number of unfavorable conditions in $\{r, w, m, v\}$, i.e., the number of values $1$ in $\{r, w, m, v\}$. This means that the AV receives a high reward if the probability of an unexpected event to occur is high, e.g., the AV is under very unfavorable conditions, and if the unexpected event is detected. This definition is to encourage the AV to use the radar mode when the environment conditions are unfavorable. In summary, the immediate reward can be defined as follows:
\begin{equation}
\label{eq:reward}
	r_t = 
	\begin{cases}
	       +r_1 , &\text{if  } \mbox{} a_t = 0, c = 0, \text{ given } \ominus, \\	
		+r_2 , &\text{if  } \mbox{} a_t = 0, c = 1, \text{ given } \ominus, \\	
		  -r_3, &\text{if  } \mbox{} a_t = 0, \text{ given }\oplus,         \\
		+r_4(b+1) , &\text{if  } \mbox{} a_t = 1, \text{ given } \oplus, \\
		0 , &\text{if } \mbox{} a_t = 1, \text{ given } \ominus.
	\end{cases}
\end{equation}
where $b$ is the number of values of $1$ in the set $\{r, w, m, v\}$. 
Note that the probability of an unexpected event to occur given $\{r, w, m, v\}$, $P(\oplus)$, is defined in (\ref{eq:unexpected-ev-prob}).

In this paper, we aim to find the optimal policy for the AV, denoted by $\pi^*$, to maximize its long-term discounted cumulative reward, i.e., discounted return, as defined by
\begin{equation}
\max_{\pi} \ G(\pi) = \mathbb{E} \lbrace \sum_{t=0}^{T} {\gamma^t r_{t+1}(\pi)} \rbrace,
\end{equation}
where $G(\pi)$ is the expected discounted return under the policy $\pi$, $r_{t+1}(\pi)$ is the immediate reward under policy $\pi$ at time step $t+1$, $T$ is the time horizon, and $\gamma \in (0, 1)$, is the discount factor. The optimal policy $\pi^*$ will allow the AV to make optimal decisions at any state $s_t$, i.e., $a^*_t = \pi^*(s_t)$.

To find the optimal policy for the AV, standard Q-learning~\cite{watkins1992} can be adopted by estimating Q-values of all state-action pairs, i.e., $Q(s,a)$. The Q-values are iteratively updated in a Q-table, and thus the Q-learning suffers the large state space problem. Therefore, we propose to use the DRL with DQN to quickly find the optimal policy.

\section{Deep Reinforcement Learning Algorithm}
The DQN algorithm uses a deep neural network, called Q-network, with weights $\boldsymbol{\theta}$ to derive an approximate value of $Q^*(s,a)$. The input of the Q-network is one of the states of the AV, and the output includes Q-values $Q(s,a,\boldsymbol{\theta}$) of all possible actions. The approximate Q-values allow the AV to map its state to an optimal action. For this, the Q-network needs to be trained to update the weights $\mathbf{\boldsymbol{\theta}}$ as follows. 

At the beginning of iteration $t$, given state $s_t \in \mathcal{S}$, the AV obtains the Q-values $Q(s, \cdot,\boldsymbol{\theta})$ for all possible actions $a$. The AV then takes an action $a_t$ according to the $\epsilon$-greedy policy~\cite{mnih2015} and observes the reward $r_t$ and next state $s_{t+1}$. The AV stores the transition $m_t = (s_t, a_t, r_t, s_{t+1})$ to a replay memory $\mathcal{M}$. Then, the AV randomly samples a mini-batch of the transitions from $\mathcal{M}$ to update $\boldsymbol{\theta}$ as follows:
\begin{equation}
\label{eq:dqn-update}
	\boldsymbol{\theta_{t+1}} = \boldsymbol{\theta_t} + \alpha \left[ y_t - Q(s_t, a_t; \boldsymbol{\theta_t}) \right] \nabla Q(s_t, a_t, \boldsymbol{\theta_t}),
\end{equation}
where $\alpha$ is the learning rate, $\nabla Q(s_t, a_t, \boldsymbol{\theta_t})$ is the gradient of $Q(s_t, a_t, \boldsymbol{\theta_t})$ with respect to the online network weights $\boldsymbol{\theta_t}$, and $y_t$ is the target value. $y_t$ is defined as $y_t = r_t + \gamma \max_a Q(s_{t+1},a; \boldsymbol{\theta}^-_t)$, where $\gamma$ is the discount factor, and $\boldsymbol{\theta}^-_t$ are the target network weights that are copied periodically from the online network weights. The above steps are repeated in iteration $t+1$ to update the weights $\boldsymbol{\theta}$. Note that the training process is considered to be an episodic task, and the algorithm converges when the cumulative reward is stable over episodes.
\vspace{-0.3cm}
\section{Performance evaluation}

\begin{figure}[]
	\centering
	\begin{minipage}[t]{0.24\textwidth}
	 \includegraphics[width=1.1\textwidth]{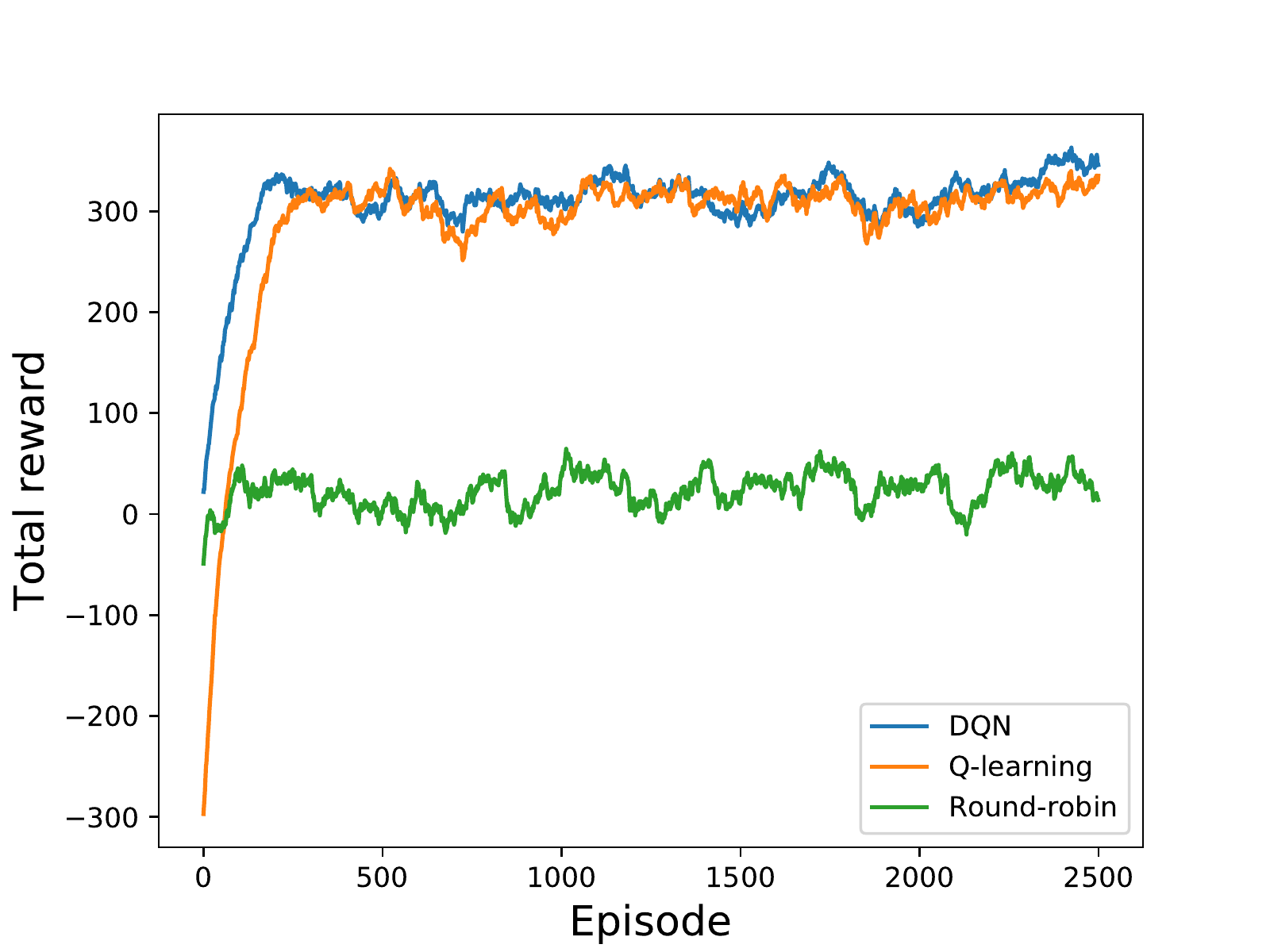}
	\subcaption{}
	\label{subfig:reward}
	\end{minipage}
	\begin{minipage}[t]{0.24\textwidth}
		\includegraphics[width=1.1\textwidth]{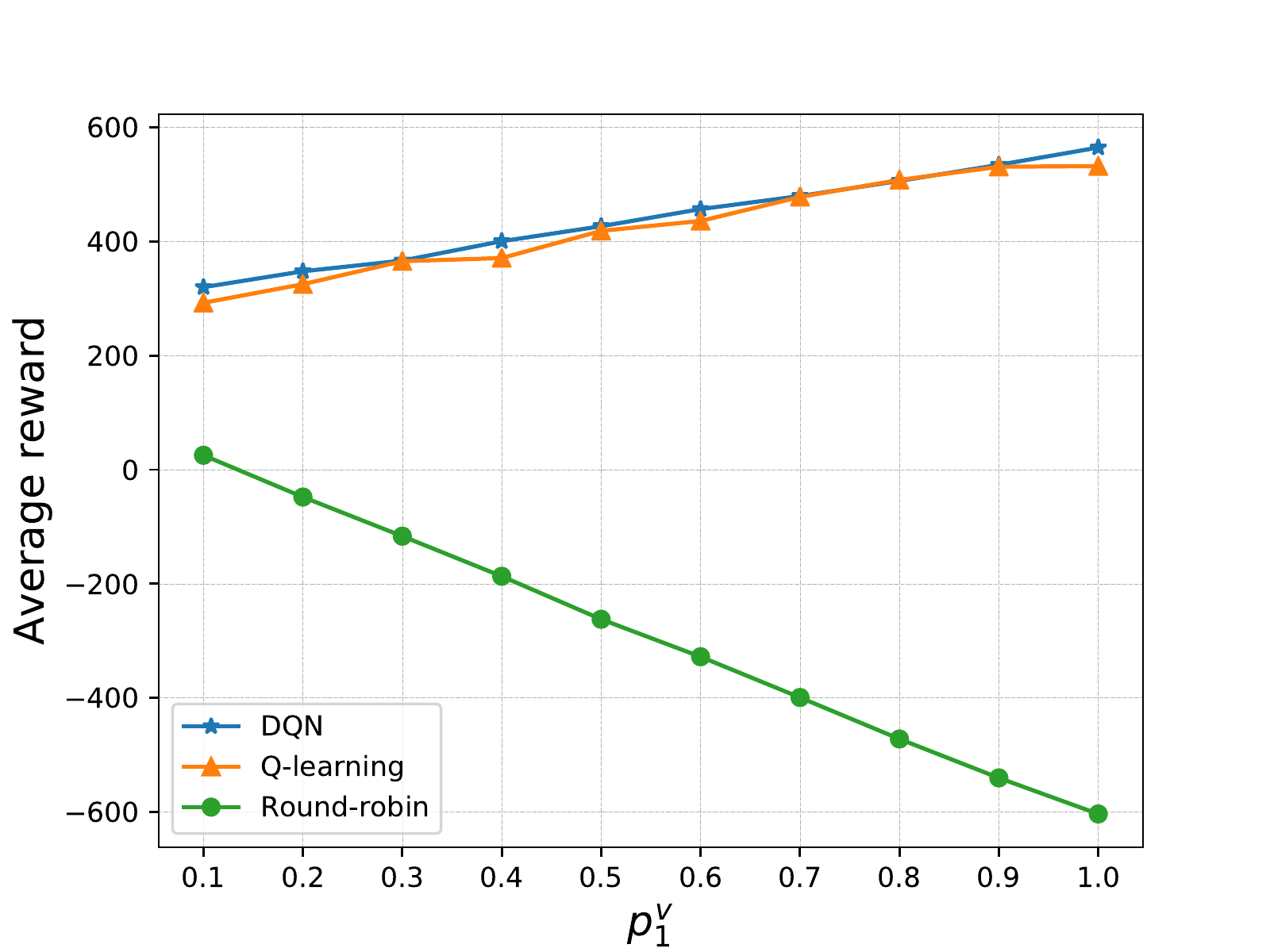}
		\subcaption{}
		\label{subfig:reward-vs-speed}
	\end{minipage}
		  \vspace{-0.2cm}
	\begin{minipage}[t]{0.24\textwidth}
		\includegraphics[width=1.1\textwidth]{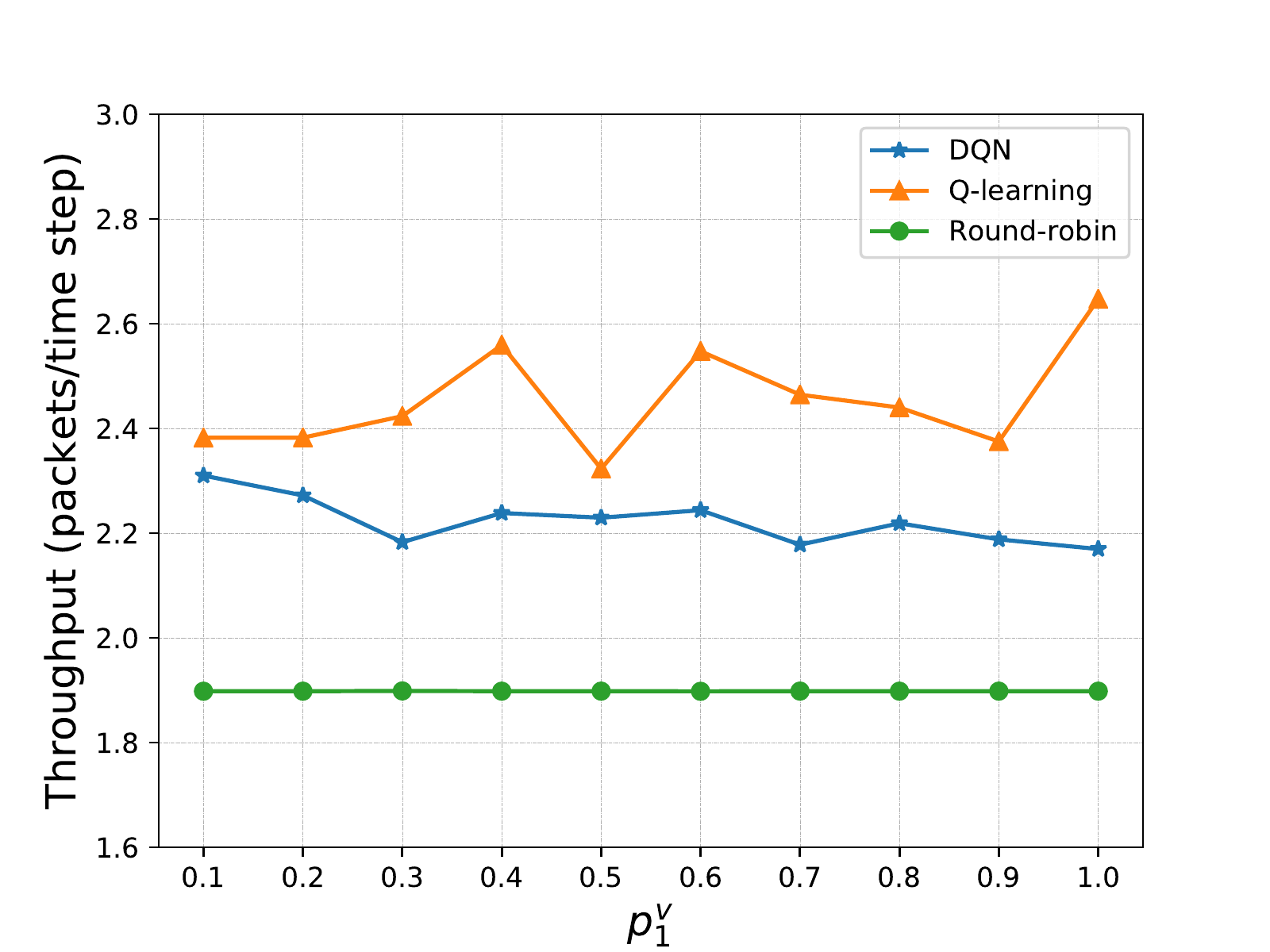}
		\subcaption{}
		\label{subfig:throughput-vs-speed}
	\end{minipage}
	\begin{minipage}[t]{0.24\textwidth}
		\includegraphics[width=1.1\textwidth]{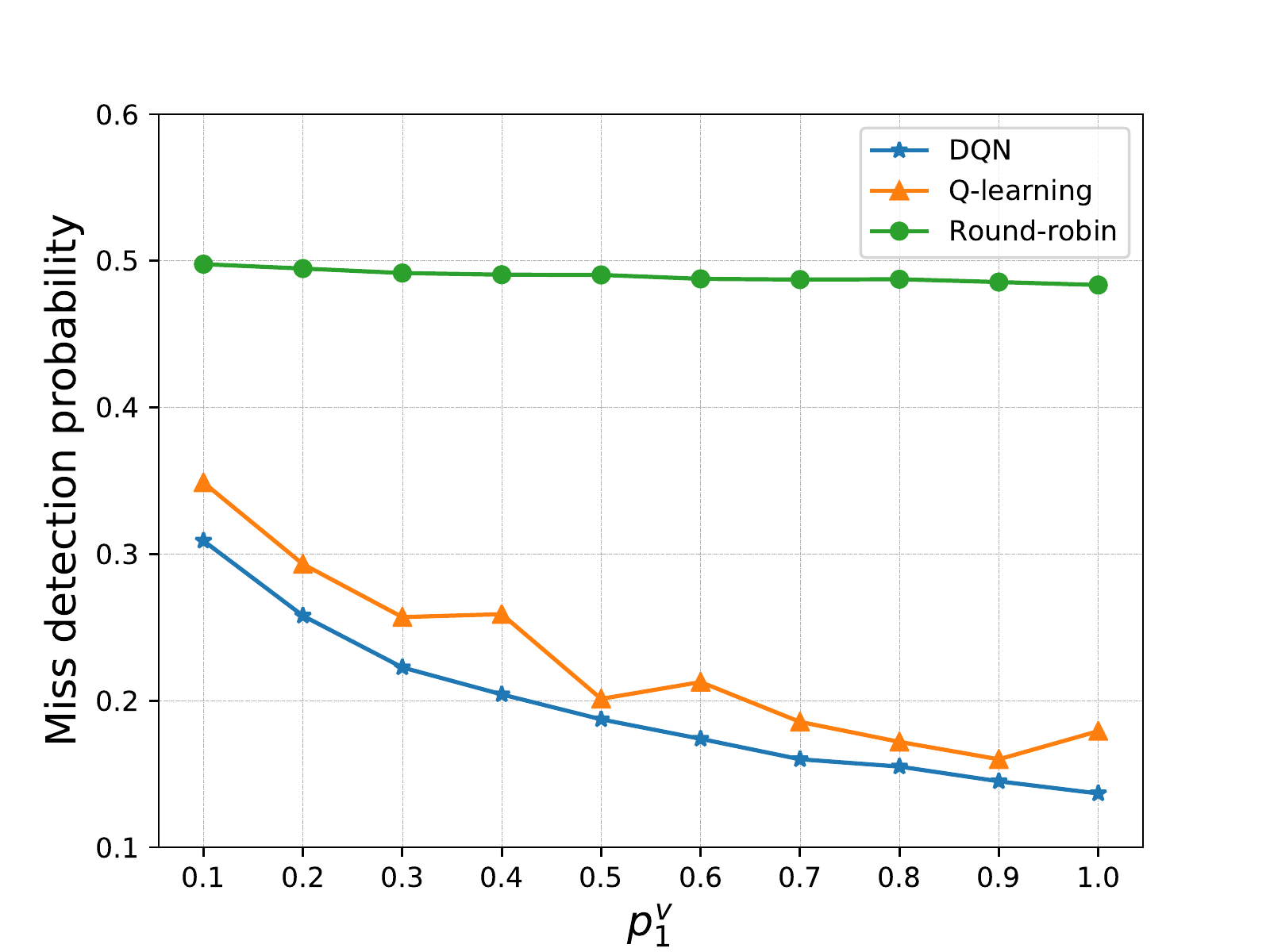}
		\subcaption{}
		\label{subfig:miss-detection-vs-speed}
	\end{minipage}
	\caption{(a) Total reward vs. episode; (b) average reward, (c) throughput, and (d) miss detection probability vs. $p_1^v$.}
	\label{fig:speed-varied}
	  \vspace{-0.5cm}
\end{figure}

\subsection{Experiment Setup}
For the comparison purpose, the capacity of the data queue is set to $D = 10$ packets, and the arrival packets follow a Poisson distribution with an average arrival rate of $\lambda_d = 1$ packet/time step.
If the channel state is good, i.e., $c=0$, the AV can transmit $\nu_1 = 4$ packets, if the channel state is bad, i.e., $c = 1$, the AV can transmit $\nu_2 = 2$ packets.
We assume that the probability that the channel is at the bad state is $p_c = 0.1$, and the probability that the channel is at the good state is $1-p_c$. For the reward values, to minimize the miss detection probability, the value of $r_3$ should be much higher than other values, i.e., $r_1$, $r_2$, and $r_4$. In particular, we set the values $(r_1, r_2, r_3, r_4)$ to be $(2, 1, 50, 5)$. The values of $p_0^v$ and $p_1^v$ are taken from~\cite{kloeden2001} in which if the AV's speed exceeds $60$ km/h, the AV's speed is high and otherwise the AV's speed is low. Specifically, the values $p_0^v$ and $p_1^v$ are set to be $0.005$ and $0.1$, respectively. Rain can be considered to be a common unfavorable weather state, and thus the values of $p_1^w$ and $p_0^w$  can be taken from~\cite{ustransport} in which $p_1^w = 0.046$ and $p_0^w = 0.005$. 
The parameters of the DQN scheme are set as follows. The neural network used in DQN is a Multilayer Perceptron with $1$ input layer, $2$ hidden layers and $1$ output layer. The input layer contains $6$ units which correspond to the number of dimensions of the state space. The output layer contains $2$ units corresponding to the number of dimensions of the action space of the AV. The DQN and the environment for the AV are implemented by using Keras library and OpenAI Gym environment, respectively. To evaluate the DQN scheme, we introduce the Q-learning~\cite{watkins1992} and  Round-robin scheme, i.e., the AV switches back and forth between the radar mode and the communication mode, as baseline schemes.

\subsection{Simulation Results}
\label{subsec:results}
We first compare the total rewards obtained by the schemes. As shown in Fig.~\ref{fig:speed-varied}(a), the total rewards obtained by the DQN and Q-learning are much higher than that of the Round-robin. Furthermore, the DQN and Q-learning converge to the same reward. 
However, the convergence speed of the DQN is much faster than that of the Q-learning. In particular, the DQN requires $170$ episodes to approach the optimal value, while the Q-learning scheme requires $280$ episodes.
The reason is that the DQN updates multiple Q-values in a mini-batch at each training iteration~\cite{mnih2015}, while Q-learning performs only one Q-values update at each training iteration~\cite{watkins1992}. As a result, the convergence rate of the Q-learning is usually much lower than that of the DQN, especially for the large state/action spaces~\cite{mnih2015}.



Next, we evaluate the DQN scheme by varying the environmental factors. Without loss of generality, we evaluate the proposed scheme when the probability to occur an unexpected event given the high speed of the AV, $p_1^v$, varies from $0.1$ to $1$. As shown in Fig.~\ref{fig:speed-varied}(b), as $p_1^v$ increases, the average reward obtained by the Round-robin scheme decreases, while those obtained by the DQN and Q-learning schemes increase. The reason can be explained as follows. With the Round-robin scheme, the radar mode is chosen according to a fixed policy, meaning that the radar mode may not be frequently used even if the occurrence probability of an unexpected event is high. Thus, the AV may receive high penalties that results in a decrease of the average reward. With the DQN and Q-learning schemes, the AV uses the radar mode more frequently as $p_1^v$ increases to minimize the penalties. As a result, the DQN and Q-learning schemes can achieve higher average rewards compared with that of the Round-robin scheme. 

Following the optimal policy, the DQN and Q-learning can significantly outperform the Round-robin in terms of throughput (see Fig.~\ref{fig:speed-varied}(c)) and miss detection probability (see Fig.~\ref{fig:speed-varied}(d)). As shown in Fig.~\ref{fig:speed-varied}(d), the miss detection probabilities obtained by the DQN and Q-learning decrease as $p_1^v$ increases. The reason is that the optimal policies obtained by the DQN and Q-learning enable the AV to select the radar mode more frequently as unexpected events are likely to occur. Thus, the AV can detect more unexpected events and reduce the miss detection probability. Note that our simulation results presented in this section are especially useful to design key parameters for real AV systems to ensure the safety for the users. In particular, given the current simulation setting $(r_1, r_2, r_3, r_4) = (2, 1, 50, 5)$, the AV can achieve a miss detection probability ranging from $0.15$ to $0.3$. We can further reduce the miss detection probability of the AV to meet its requirement by increasing the reward when the AV selects the radar mode, e.g., increasing $r_4$ from $5$ to $50$ or $100$.

\section{Conclusion}
In this paper, we have proposed the iRDRC system which enables the AV to optimize the radar mode and communication mode selection automatically in a real-time manner. To deal with the uncertainty of the environment, we have formulated the optimization problem based on the MDP framework and developed the DQN algorithm to obtain the optimal policy. The results show that the proposed system can simultaneously maximize the data throughput and minimize miss detection probability. 
In our future work, continuous actions and cooperation between the AVs in V2I networks can also be considered. 


\end{document}